\documentclass[aps, 10pt, pra, amsmath, longbibliography, twocolumn, superscriptaddress, floatfix]{revtex4-1}
\usepackage[plainpages=false,colorlinks=true,linkcolor=blue, citecolor=blue, urlcolor=blue]{hyperref}
\usepackage[english]{babel}
\usepackage{graphicx}
\usepackage[charter]{mathdesign}

\newcommand{\bse}{\begin{subequations}}
\newcommand{\ese}{\end{subequations}}

\newcommand{\beq}{\begin{equation}}
\newcommand{\eeq}{\end{equation}}
\newcommand{\bea}{\begin{eqnarray}}
\newcommand{\eea}{\end{eqnarray}}
\newcommand{\ve}{\varepsilon}
\newcommand{\up}{\uparrow}
\newcommand{\down}{\downarrow}

\newcommand{\bk}{{\bf k}}

\newcommand{\bwt}{\begin{widetext}}
\newcommand{\ewt}{\end{widetext}}

\newcommand{\er}{\eqref}
\newcommand{\itx}{\textit}

\newcommand\kv{\mathbf{k}}

\begin{document}

\title{Cluster dynamical mean field study of intra-unit-cell charge nematicity in hole-doped cuprates} 
\author{Abhishek Kumar}
\author{David S\'en\'echal}
\author{A.-M. S. Tremblay}
\affiliation{D\'epartement de physique and Institut Quantique, Universit\'e de Sherbrooke, Sherbrooke, Qu\'ebec, Canada J1K 2R1}

\begin{abstract}
Recent scanning-tunneling microscopy on hole-doped Bi$_2$Sr$_2$CaCu$_2$O$_8$, one of the materials of the cuprate family, finds a long-range ordered spontaneous splitting of the energy levels of oxygen orbitals inside the CuO$_2$ unit cells [S. Wang et al., \itx{Nat. Mat.} 23, 492-498 (2024)]. 
This spontaneous intra-unit-cell orbital ordering, also known as electronic nematicity, breaks $C_4$ symmetry and is thought to arise from the Coulomb interaction (denoted by $V_{pp}$) between oxygen $p_x$ and $p_y$ electrons.
In this work, we study the spontaneous emergence of electronic nematicity within the three-band Hubbard (aka the Emery-VSA model), using
cluster dynamical mean field theory.
This method incorporates short-range electronic correlations and gives us access to the density of states, a quantity that is directly probed in experiments.
We argue that there is a delicate competition between $V_{pp}$ and $V_{pd}$ (the latter being the Coulomb interaction between copper $d_{x^2-y^2}$ and oxygen $p_{x,y}$ electrons) that must be taken into account in order to find a Zhang-Rice singlet band well-resolved from the upper Hubbard band, and a splitting of the charge-transfer band (one of the signatures of charge nematicity) by roughly 50 meV, as observed recently. 
\end{abstract}

\maketitle

\section{Introduction}

One of the most enigmatic phases of hole-doped cuprates is charge nematicity~\cite{fradkin:rmp}.
This long-ranged charge order has been investigated quite frequently in the past two decades~\cite{Nakata:2021, Auvray:2019, Mangin:2017, Achkar:2016, Gael:2015, Xie:2022, Wu:2018, Wu:2020, Murayama:2019, Daou:2010, Mook:2008, Fauque:2006, Li:2011, Li:2008, Fujita:2014, Zhao:2017, Sato:2017}. 
It manifests itself as the spontaneous breaking, by the electronic structure, of the four-fold rotational symmetry of the CuO$_2$ unit cell~\cite{Daou:2010} common to all cuprates.

Even after numerous experimental studies in the past 15 years, a microscopic mechanism has been experimentally established only recently~\cite{wang2023_nema}:
using sublattice-resolved spectroscopic imaging scanning tunneling microscopy (STM) on Bi$_2$Sr$_2$CaCu$_2$O$_{8+\delta}$~\cite{wang2023_nema}, a splitting in the charge-transfer energy (CTE)~\cite{lawler:2010, mesaros:2011, fujita:pnas} of $p_x$ and $p_y$ oxygen orbitals within the CuO$_2$ unit-cell has been measured, which is a direct evidence of intra-unit-cell (IUC) charge nematicity in hole-doped cuprates.
Several theoretical works~\cite{kivelson:2004, fischer:2011, atkinson:2013, maier:2014, fischer:2014, hiroshi:2018, shiwei:2018, yamase:2021, Zegrodnik:2020} studied the problem using the three-band Emery-VSA (Varma-Schmitt-Rink-Abrahams) model~\cite{emery:1987, varma:1987}.
Strong coupling  approaches \cite{kivelson:2004} and  self-consistent mean-field theory (within the Hartree approximation only)~\cite{fischer:2011} have shown how such a long-range orbital ordering can arise from the Coulomb repulsion $V_{pp}$
between electrons located on the two different oxygen orbitals ($p_x$ and $p_y$) of that model.
The Emery model with $V_{pp}$ interaction has also been studied with weak-coupling perturbation theory~\cite{atkinson:2013, maier:2014} and with a strong-coupling perturbation expansion to look for charge nematic fluctuations~\cite{fischer:2014}.
It has also been claimed that nematicity can arise even when $V_{pp}=0$~\cite{hiroshi:2018, shiwei:2018, yamase:2021, Zegrodnik:2020}.

So far, however, the interplay of $V_{pp}$ with other Coulomb interaction terms, such as $V_{pd}$, the nearest-neighbor interaction between Cu $d_{x^2-y^2}$ and O $p_{x, y}$ electrons, and $U_p$, the O on-site interaction, has not been investigated.
Our contribution is twofold. 
First, we show that the competition between $V_{pp}$ and $V_{pd}$ must be considered in order to establish small IUC nematicity close to half-filling, as observed in recent STM experiment \cite{wang2023_nema}.
Second, we show that the calculation of partial density of states (DOS) for the Cu and O orbitals in the interacting model is necessary, not only to make contact with the experiment \cite{wang2023_nema}, 
but also to reject some values of the interaction parameter set \{$U_d$, $U_p$, $V_{pp}$, $V_{pd}$\} that give incorrect DOS, despite finding the same charge nematicity. 

Since cuprates are in an intermediate coupling regime where neither band structure nor interactions dominate,  we need a method that is valid in this difficult case where the DOS must exhibit a charge-transfer band, a Zhang-Rice singlet band and an upper Hubbard band.  
Cluster dynamical mean field theory (CDMFT)~\cite{kotliar:2001, maier:rmp, KotliarRMP:2006, LTP:2006, senechal:2012, dionne:2023} satisfies these requirements. 
We analyze the same three-band Emery-VSA model as in  previous works \cite{kivelson:2004, fischer:2011, atkinson:2013, fischer:2014, maier:2014, hiroshi:2018, shiwei:2018, yamase:2021}. 
 
We find that the system develops IUC charge nematic order near half-filling when $V_{pp}$ is large enough. 
However, we also find that $V_{pd}$ pushes the onset of the nematic transition at large doping for large-enough $V_{pd}$. 
For small $V_{pd}$, on the other hand, the Zhang-Rice singlet band (ZRSB)~\cite{mai:2021, unger:1993, devereaux:2013}, one of the characteristic features of hole-doped cuprates (with mixed Cu-O character), hybridizes with the upper Hubbard band (UHB) of Cu, contrary to observations. 
We thus argue that the calculation of the DOS is crucial to determine the optimal value of $V_{pd}$ and also of other parameters, such as the on-site oxygen repulsion $U_p$. 
We end with a comparison of our CDMFT results with those obtained from static mean-field theory (MFT) within both the Hartree and Hartree-Fock approximations. 
This will clearly illustrate that similar charge nematicity with these two methods give drastically different DOS, allowing us to illustrate the importance of using CDMFT.

\section{Model and order parameter}
\label{sec:model_method}

\subsection{Model Hamiltonian}
\label{sec:model}

The noninteracting part of the Emery model which describes hopping of electrons in the CuO$_2$ plane \cite{fischer:2011} is expressed as follows in real space:
\beq
\label{non-int}
\begin{split}
\mathcal{H}_0 &= \,\, \mathcal{H}_\text{kin} - \mathcal{H}_\mu, \\
\mathcal{H}_\text{kin} &= t_{pd} \sum_{i, s} \sum_{\nu=\pm x, \pm y} (\hat{d}_{i, s}^\dagger \hat{p}_{i+\nu/2, s} + \text{H.c.}) \\
&+ t_{pp} \sum_{i, s} \sum_{\nu=\pm x} \sum_{\nu'=\pm y}
(\hat{p}_{i+\nu/2, s}^\dagger \hat{p}_{i+\nu'/2, s} + \text{H.c.}) \\
\mathcal{H}_\mu &= \mu \sum_{i, s} \hat{n}_{i, s}^d + \frac{1}{2} (\mu-\Delta) \sum_{i, s} \sum_{\nu=\pm x, \pm y} \hat{n}_{i+\nu/2, s}^p,
\end{split}
\eeq
where $t_{pd}$ is Cu-O hopping integral and, $t_{pp}$ and $t_{pp}'$ are O-O nearest-neighbor and next-nearest-neighbor hopping integrals in the CuO$_2$ unit cell, respectively. 
Here, $\hat{d}_{i, s}^\dagger$ creates an electron with spin $s$ in the copper $d_{x^2-y^2}$ orbital at site $i$ and $\hat{p}_{i+\nu/2, s}^\dagger$ creates an electron with spin $s$ on the oxygen $p_\nu$ orbital at the site $i+\nu/2$ for $\nu = \pm x, \pm y$, with $\langle \nu, \nu' \rangle$ pointing to neighboring oxygen sites.
The Cu atoms at sites $i$ form a square lattice with unit vectors $\hat{x}$ and $\hat{y}$, lattice spacing unity, and total number of lattice sites $N$. 
The chemical potential $\mu$ and the bare charge-transfer energy $\Delta \equiv \Delta_{\text{O}_{x,y}} - \Delta_\text{Cu}$ control the total and relative electron densities of Cu and O, whose number operators are defined as $\hat{n}_{i, s}^d$ and $\hat{n}_{i+\nu/2, s}^p$, respectively.

\begin{figure*}
\includegraphics[width=1\linewidth]{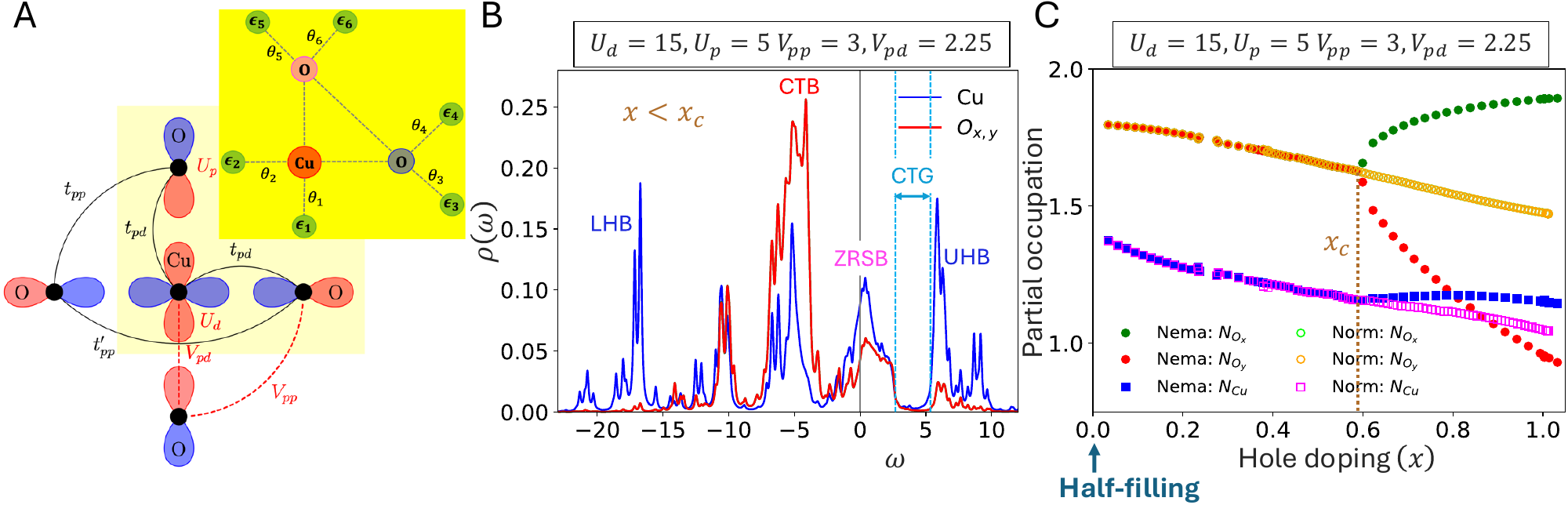}
\caption{\label{fig:comb}
(\itx{A}) Schematic view of the Emery model showing hopping and interaction parameters appearing in Eqs~\ref{non-int} and \ref{int}. 
Inset: the 3-site impurity model, consisting of one of each orbital ($d$, $p_x$, $p_y$) in the unit cell, along with uncorrelated bath orbitals -- green circles) used for the CDMFT computation. 
(\itx{B}) Partial densities of states (DOS) of the interacting model for $x<x_c$ when the system is in the normal (or non-nematic) phase. 
Here, LHB and UHB stand for lower and upper Hubbard band, CTG and CTB for charge-transfer gap and charge-transfer band. and ZRSB for Zhang-Rice singlet band. 
The Grey vertical line denotes the location of the chemical potential ($\mu$), which, at finite hole doping, lies in the ZRSB. 
The system is in the normal phase, so the $p_x$ and $p_y$ orbitals (red curve) are degenerate. 
(\itx{C}) The occupation of Cu and planar O ($p_x$ and $p_y$) orbitals vs hole doping ($x$) in the CuO$_2$ unit cell.
Each orbital holds at most 2 electrons, so at full-filling there are 6 electrons in the unit-cell. 
In the conventional hole picture, the fully filled, half-filled and empty unit-cell correspond to $x=-1$, $x=0$ and $x=5$, respectively. 
The open and solid symbols correspond to normal-state and nematic-state computations, respectively, for band parameters given by Eq.~\ref{param} and interaction parameters as printed in the figure. 
In the normal phase ($x<x_c$), the open and solid symbols coincide exactly, while in the nematic phase ($x>x_c$) a charge imbalance between $p_x$ (green) and $p_y$ (red) orbitals develops beyond $x=x_c$. 
}
\end{figure*}

The interaction part of the Emery model includes on-site interactions with strengths $U_d$ and $U_p$ on Cu and O orbitals respectively, as well as extended interactions, $V_{pd}$ (between nearest neighbors Cu and O) and $V_{pp}$ (between neighboring oxygens) \cite{fischer:2011}:
\beq
\label{int}
\begin{split}
\mathcal{H}_\text{int} =& U_d \sum_i \hat{n}_{i \up}^d \hat{n}_{i \down}^d + \frac{U_p}{2} \sum_{i, \nu=\pm x, \pm y} \hat{n}_{i+\nu/2, \up}^p \hat{n}_{i+\nu/2, \down}^p \\
& + V_{pd} \sum_{i} \sum_{s, s'} \sum_{\nu=\pm x, \pm y}\hat{n}_{i, s}^d \hat{n}_{i+\nu/2, s'}^p \\
& + V_{pp} \sum_i \sum_{s, s'} \sum_{\nu=\pm x} \sum_{\nu'=\pm y} \hat{n}_{i+\nu/2, s}^p \hat{n}_{i+\nu'/2, s'}^p.
\end{split}
\eeq
A pictorial representation of the CuO$_2$ unit cell indicating various hopping and interactions, as given by Eqs.~\ref{non-int} and \ref{int}, is provided in Fig.~\ref{fig:comb}\itx{A}.

The Emery model describes a charge-transfer insulator at large values of $U_d$ and at a filling of one hole per unit cell (we refer to this as the undoped state, or as half-filling). 
Such an insulating state is realized by tuning $U_d$: as $U_d$ increases, the Cu band splits into lower and upper Hubbard bands (LHB and UHB) and the UHB is eventually pushed beyond the oxygen-dominant band (near $\Delta$), leading to an insulating gap between the latter two,  referred to as the charge-transfer gap (CTG)~\cite{zaanen:1985}. 
The central, oxygen-dominated band is called the charge-transfer band (CTB). 
On doping the charge-transfer insulator, the holes primarily go into the oxygen orbitals and another band appears at the Fermi level, referred to as the Zhang-Rice singlet band (ZRSB)~\cite{mai:2021, unger:1993, devereaux:2013}.

The density of states (DOS) of the interacting model exhibiting all the above features is shown in Fig.~\ref{fig:comb}\itx{B} at some doping $x$, for band parameters given by 
\beq
\label{param}
\Delta \equiv \Delta_{\text{O}_{x,y}}-\Delta_\text{Cu} = -5.8\qquad,\qquad t_{pd}=2.1~~.
\eeq
Here, all energies are given in units of $t_{pp}\sim 0.65$ eV \cite{Weber:2012, AMT:pnas} and
$\Delta_\text{Cu}=0$ was used as the energy reference.
In Fig.~\ref{fig:comb}\itx{B}, the system is in the normal phase for $x<x_c$, where $x_c$ is some critical hole doping above which the systems undergoes a transition to the nematic phase. 
In the normal phase, the oxygen $p_x$ and $p_y$ orbitals are degenerate, as shown by the red band in Fig.~\ref{fig:comb}\itx{B}.
The degeneracy is lifted for $x>x_c$ when the system enters the nematic phase.

\subsection{Charge nematic order}
\label{sec:order}

The IUC charge nematic order ($\eta$) is defined as a spontaneous imbalance between the densities of the $p_x$ and $p_y$ oxygen orbitals of the Emery model:
\beq
\label{nem_order}
\eta \equiv |(n_{p_x\up} + n_{p_x\down}) - (n_{p_y\up} + n_{p_y\down})| = |n_{p_x} - n_{p_y}|.
\eeq
We do not allow any magnetization of the O sites ($n_{p_{x,y}\up} = n_{p_{x,y}\down}$), hence the last equality above. 
As an illustration, the density imbalance of $p_x$ and $p_y$ electrons is shown in Fig.~\ref{fig:comb}\itx{C}, where the green ($p_x$) and red ($p_y$) curves bifurcate at $x=x_c$, indicating the onset of the nematic transition. 
Of course, the critical doping $x_c$ depends on the choice of interaction parameters.

In the normal phase, the occupation of Cu (squares) is small compared to that of O (circles), which is expected since the UHB of Cu lies above the Fermi level (see Fig.~\ref{fig:comb}\itx{B}). 

In the nematic phase ($x>x_c$) the holes primarily go to the O-band upon doping: while the slope of blue curve (Cu) is almost zero, that of green and red curves (O$_x$ and O$_y$) varies significantly when $x>x_c$. 
This is in agreement with experiments \cite{tranquada:1987, emery:1988, chen:prl, gauquelin:2014} 
and recent CDMFT studies \cite{Fratino_Semon_Sordi_Tremblay_2016, AMT:pnas} in the context of superconductivity. 

We performed CDMFT computations at zero temperature, using an exact diagonalization (ED) impurity solver, for the 3-site impurity model shown in Fig.~\ref{fig:comb}\itx{A}(inset) to obtain Figs.~\ref{fig:comb}\itx{B} and \ref{fig:comb}\itx{C}.  
Clearly, the results shown in Fig.~\ref{fig:comb}\itx{C}, though obtained from actual computation, do not depict the most favorable scenario for experiments, as the onset of nematic transition occurs at high doping ($x\approx0.6$). 
So, our goal is to find a realistic set of interaction parameters for which small nematicity is found at low doping.  
For more details on the CDMFT procedure and the cluster model, see Appendix~\ref{Sec:methods}.

\section{Results}
In order to elucidate the origin of IUC charge nematicity in hole-doped cuprates, we computed the nematic order parameter for various interaction parameters as a function of hole doping ($x$), with band parameters given by Eq.~\ref{param}.  
Even after setting the band parameters, the model still has a large parameter space, \{$U_d$, $U_p$, $V_{pd}$, $V_{pp}$\} to play with; therefore mapping out a single phase diagram is impossible. 
Hence we compute the nematic order by varying one interaction parameter at a time, while keeping the others fixed at some realistic value. 
This we sequentially do for $V_{pp}$, $V_{pd}$ and $U_p$, assuming fixed $U_d$, such that $U_d \gg U_p, V_{pp}, V_{pd}$.
We also compute the density of states (DOS), which is one of the direct ways to verify the signature of a quantum nematic phase transition. 

\subsection{Charge nematicity induced by first-neighbor interactions}
\label{sec:orbit}

Early works \cite{kivelson:2004, fischer:2011} have focused on the role of $V_{pp}$ on charge nematicity.
However, the Cu-O interaction $V_{pd}$ has generally been neglected, which is a priori unjustified. 
In this section, we argue that the interplay of these two interactions is crucial for charge nematicity.

\subsubsection{Effect of the O-O repulsion}
\label{role_Vpp}
\begin{figure*}
\includegraphics[width=1\linewidth]{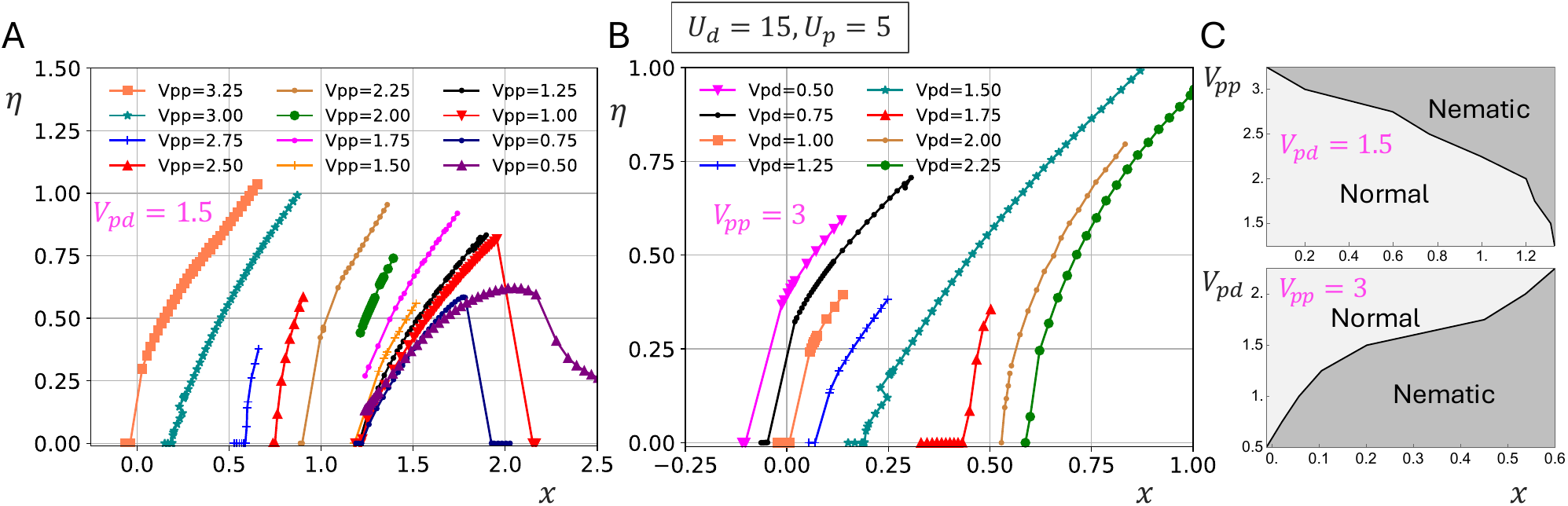}
\caption{\label{fig:phase} 
Nematic order parameter ($\eta$) as a function of hole doping ($x$) and phase diagrams for $V_{pp}$ and $V_{pd}$ vs $x$. 
(\textit{A}) CDMFT data showing the effect of $V_{pp}$ on IUC charge nematicity. 
For fixed $U_d$, $U_p$ and $V_{pd}$, increasing $V_{pp}$ shifts the onset of nematic transition towards low doping, thus favoring nematicity.  
(\textit{B}) The effect of $V_{pd}$ on nematicity. This time, for fixed $V_{pp}$, increasing $V_{pd}$ shifts the onset of nematicity towards higher doping. 
(\textit{C}) $V_{pp}$ vs $x$ (top) and $V_{pd}$ vs $x$ (bottom) phase diagrams, from data in panels (\textit{A}) and (\textit{B}).
}
\end{figure*}

We consider realistic values of $U_d$, $U_p$ and $V_{pd}$ \cite{hybertsen:1989, mcmahan:1988, stechel:1988} and perform CDMFT computation for different $V_{pp}$'s to understand its effect on charge nematicity. 
For a particular $V_{pp}$, we scan the chemical potential (which controls the electronic density) and find the onset of the nematic transition. 
The results are shown in Fig.~\ref{fig:phase}\itx{A}, and the corresponding phase diagram in Fig.~\ref{fig:phase}\itx{C}(top).

Fig.~\ref{fig:phase}\itx{A} depicts the general behavior of the nematic order parameter $\eta$ as a function of hole doping $x$. 
We observe that, at a given $V_{pp}$, as doping is increased, the system undergoes a first-order transition to the nematic phase. 
A large $V_{pp}$ favors nematicity, since the onset of the nematic transition shifts to low doping as $V_{pp}$ increases.
As per this trend, nematicity disappears at all doping at small enough $V_{pp}$: in Fig.~\ref{fig:phase}\itx{A}, for given $U_d$, $U_p$ and $V_{pd}$, nematicity disappears completely for $V_{pp}\lesssim 0.5$. 
Finally, for small $V_{pp}$'s, for e.g., $V_{pp}=\{0.5, 0.75, 1\}$, we observe a nematic region confined in a small window of hole doping.
A finite nematic region is expected at large $V_{pp}$'s also, but this will be achieved in a larger window of hole doping. 
Indeed, the order parameter eventually must go back down at large doping since it cannot be nonzero at zero filling. 
We already see it at small $V_{pp}$'s. 
So it is obvious that the doping range of nematicity will increase with $V_{pp}$. 
Since this is not a region of our interest (after all, the band parameters of the model depend on doping if they are derived from DFT and they would be massively different at very high doping), we do not dwell on this.

The $V_{pp}$ vs $x$ phase diagram is shown in Fig.~\ref{fig:phase}\itx{C}(top), which shows a threshold in $V_{pp}$ above which the nematicity appears.
Note that we restricted the computations to small and intermediate values of doping $x$ for reasons of economy.
No new physics is expected in the high doping region.

\subsubsection{Effect of the Cu-O repulsion}
\label{role_Vpd}
We then investigate the effect of $V_{pd}$, the Cu-O Coulomb repulsion term. 
We fix $U_d$, $U_p$ and $V_{pp}$ and compute the order parameter $\eta$ as a function of doping for a few values of $V_{pd}$. 
The CDMFT results are shown in Fig.~\ref{fig:phase}\itx{B}, and the corresponding phase diagram in Fig.~\ref{fig:phase}\itx{C}(bottom).

In Fig.~\ref{fig:phase}\itx{B}, the behavior of $\eta$ vs $x$ is similar to that of Fig.~\ref{fig:phase}\itx{A},
except for the crucial difference that whereas $V_{pp}$ favors nematicity, $V_{pd}$ hinders it.
Indeed, the nematic transition is pushed towards large doping as $V_{pd}$ increases, contrary to $V_{pp}$ in Fig.~\ref{fig:phase}\itx{A}.
This indicates an intrinsic competition between $V_{pp}$ and $V_{pd}$. 
The $V_{pd}$-doping phase diagram is shown in Fig.~\ref{fig:phase}\itx{C}(bottom).
It suggests an upper bound on $V_{pd}$ above which nematicity is impossible. 
We note that in Fig.~\ref{fig:phase}\itx{B}, the order parameter data for the lowest $V_{pd}$ ($V_{pd}=0.5$) extends in the electron-doped region (above half-filling, which is $-1<x<0$), which is plausible. 
This is likely the case also for larger $V_{pp}$ in Fig.~\ref{fig:phase}\itx{A}.

\subsection{Density of states and the optimization of the Cu-O repulsion}
\label{sec:DOS}

\begin{figure*}
\includegraphics[width=1\linewidth]{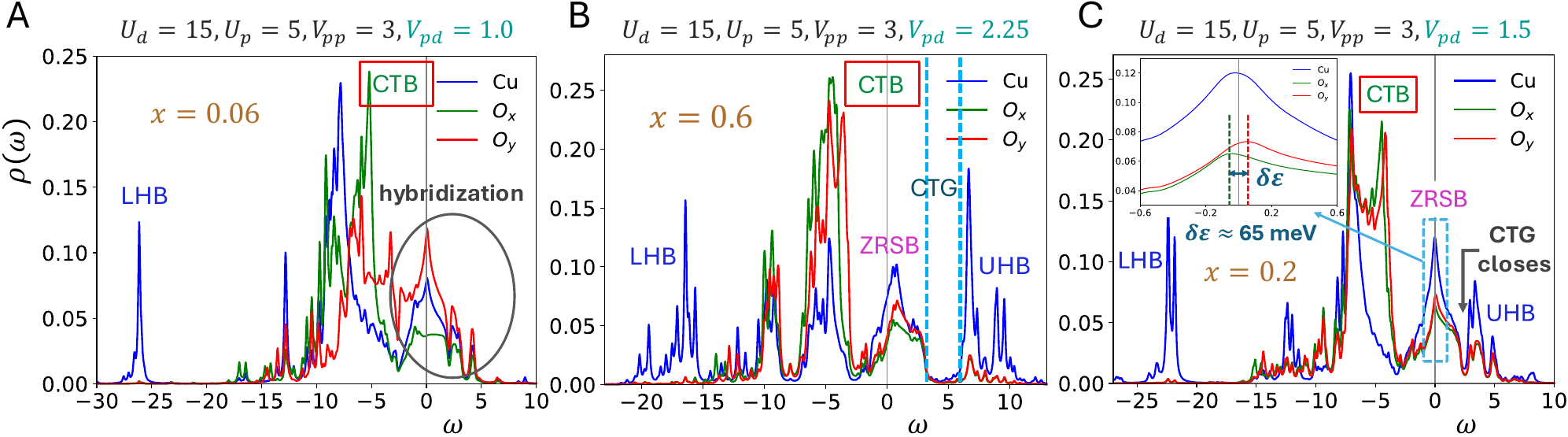}
\caption{Partial DOS in the nematic phase and optimization of $V_{pd}$. In the nematic phase, the CTB splits into two (green: $p_x$, red: $p_y$), resulting in different effective charge-transfer energies ($\ve$) of the oxygen $p_x$ and $p_y$ orbitals. 
For the values $U_d$, $U_p$ and $V_{pp}$ shown, the optimum value of $V_{pd}$ is determined from the DOS. 
(\textit{A}) At small $V_{pd}$, although the onset of the nematic transition occurs at low doping (see Fig.~\ref{fig:phase}\itx{B}), the ZRSB and the UHB hybridize strongly (encircled for clarity), contrary to observations. 
(\textit{B}) At large $V_{pd}$, the ZRSB and the UHB are separated by the CTG. However, the onset of the transition is pushed at large doping (see Fig.~\ref{fig:phase}\itx{B}).
(\textit{C}) At intermediate $V_{pd}$, the transition occurs at low-enough doping; furthermore, the ZRSB and the UHB in the DOS are well resolved (but with a very small CTG).
Inset: Near the Fermi level, the difference in effective charge-transfer energies ($\delta\ve$) is estimated to be $\delta\ve\approx 65$ meV. In all these plots, an artificial Lorentzian broadening of $0.1 t_{pp}$ is used when integrating the spectral function over momenta to compute the DOS.
}\label{fig:dos}
\end{figure*}

In Fig.~\ref{fig:comb}\itx{B} we showed the normal-state density of states (DOS), where the CTB is doubly-degenerate. 
By contrast, in the nematic phase, the CTB splits into two: there is a splitting of the energy levels ($\ve$) of the two O orbitals of the model.
This is illustrated in Fig.~\ref{fig:dos} (here, $\omega$ is in units of $t_{pp} \approx 0.65$ eV \cite{Weber:2012, AMT:pnas}), where the green and red curves represent the DOS of the $p_x$ and $p_y$ orbitals, respectively.
Measuring the energy splitting $\delta\ve$ between these two curves is one of the direct signatures of the quantum nematic phase transition in cuprates, as done recently using sublattice-resolved spectroscopic imaging scanning tunneling microscopy~\cite{wang2023_nema}.

We use the solutions obtained from CDMFT as shown in Fig.~\ref{fig:phase} to calculate the DOS. 
Since, as discussed above, a small $V_{pd}$ is favorable for nematicity (the onset of the transition is at low doping), it is tempting to consider it either very small or to ignore it altogether. 
However, we argue that a small value of $V_{pd}$ is not favorable either, not for nematicity, but because of its effect on the DOS: it leads to a strong hybridization of the upper Hubbard band (UHB) with the Zhang-Rice singlet band (ZRSB), which is not observed.
So, our goal of studying the DOS is two-fold: (i) from Fig.~\ref{fig:phase}\itx{B}, to determine the optimal value of $V_{pd}$, and (ii) at that value, from the nematic-state DOS close to the transition, extract the splitting $\delta\ve$ near the Fermi level, in order to connect with the experiment.

\subsubsection{Optimization of the Cu-O repulsion}
\label{sec:Vpd}
Panels \itx{A}, \itx{B} and \itx{C} of Fig.~\ref{fig:dos} show the DOS at small, large and optimal $V_{pd}$, respectively.
Since a small $V_{pd}$ favors nematicity, we choose $V_{pd}=1$ from Fig.~\ref{fig:phase}\itx{B} (the onset of the nematic transition is at very low doping), and calculate the DOS in the nematic state at $x=0.06$ as shown in Fig.~\ref{fig:dos}\itx{A}. 
We observe that, even though the splitting between CTBs is small, the UHB and the ZRSB hyrbidize strongly, contrary to observations and to the general behavior expected in cuprates. 
On the other hand, at large $V_{pd}=2.25$, although the DOS exhibits a clear charge-transfer gap (Fig.~\ref{fig:dos}\itx{B}), the onset of the nematic transition is pushed at large $x$, contrary to observations; Fig.~\ref{fig:dos}\itx{B} is obtained at large doping $x=0.6$. 
We conclude that an intermediate value of $V_{pd}$ is best to have the onset of the transition at low-enough doping, accompanied by the UHB and ZRSB being well resolved, albeit with a rather small charge transfer gap (CTG). 
Recent experimental observations \cite{Ruan:2016} and CDMFT computations \cite{AMT:pnas} also find that a small CTG is favorable to superconductivity, the superconduting $T_c$ reaching its maximum for small CTG \cite{Ruan:2016}. 
Recently, in Bi$_2$Sr$_2$CaCu$_2$O$_{8+x}$ \cite{wang2023_nema}, the electronic nematicity was found in the region of the phase diagram ($T$ vs $x$, where $T$ is temperature and $x$ is hole doping) where superconductivity also exists. 
A small CTG being also favorable for nematicitiy further strengthens our finding.

\subsubsection{Effective difference in charge-transfer energies between oxygen orbitals}
\label{sec:extract}
From Fig.~\ref{fig:dos}\itx{C}, the splitting ($\delta\ve$) between the $p_x$ (green) and $p_y$ (red) bands near the Fermi level is very small. 
In the inset it is blown for clarity and we find $\delta\ve \approx 65$ meV. 
We take this spliting of the oxygen DOS in the ZRSB as a measure of the experimental shift of the CTG since the edge of the upper Hubbard band is not affected by nematicity while the shift in the upper edge of the CTB in Fig.~\ref{fig:dos}\itx{C} is comparable to the shift measured in the ZRSB. 
This is close to the experimental value, reported to be $\approx 50$ meV at hole doping $x=0.17$~\cite{wang2023_nema}. 
We emphasize that the result is not unique to the interaction parameter set as chosen in Fig.~\ref{fig:dos}\itx{C}. 
In Figs.~\ref{fig:optimal} and \ref{fig:dos_SI} of Appendix.~\ref{sec:optimal}, we provide CDMFT data for the order parameter and the DOS, respectively, for other optimal interaction parameter sets, and the values of $\delta\ve$ found there roughly agree with the experiment. 

\begin{figure}\center{\includegraphics[width=0.9\linewidth]{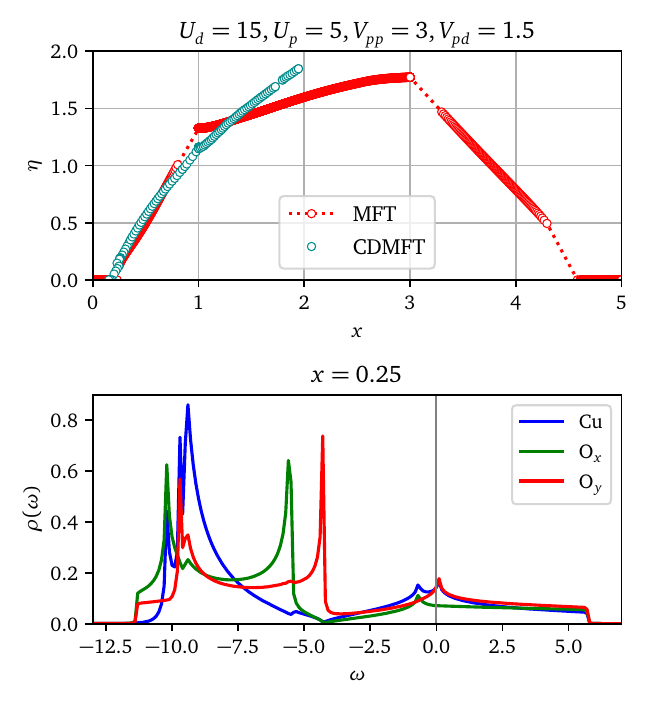}}
\caption{\label{fig:compr}
Top panel: Comparison of nematic order parameters obtained in MFT (red) and CDMFT (cyan). Bottom panel: Density of states for the mean-field solution found for the same parameters as the CDMFT solution shown in Fig.~\ref{fig:dos}\itx{C}. Note the absence of charge transfer gap.}
\end{figure}

\subsection{Effect of the oxygen on-site repulsion}
\label{sec:up}
Given the extended character of the oxygen orbitals, the value of oxygen on-site repulsion should be small. 
We take $U_p=5$, which is smaller than the on-site copper repulsion $U_d=15$. 
In Appendix.~\ref{sec:Up} (see Fig.~\ref{fig:Up_SI}), we show that a larger on-site oxygen repulsion leads to enhanced charge nematicity caused by the hybridization between the ZRSB and the UHB, as evident in Fig.~\ref{fig:dos_Up}, in contradiction with observations.
We noted a similar hybridization-enhanced nematicity with $V_{pd}$ earlier. 
We must reject such enhancements in both cases since the resulting DOS contradicts the experiment. 

\subsection{Comparison between CDMFT and static Hartree-Fock MFT}
\label{sec:comparison}
We now compare our CDMFT results with those obtained from self-consistent static MFT.
Fischer et al. have applied self-consistent MFT to the Emery model in order to probe the emergence of charge nematicity.~\cite{fischer:2011}; their analysis was limited to the Hartree approximation.
In order to compare with our CDMFT results, we also performed self-consistent MFT computations, this time
including Fock and Hartree terms (HF-MFT).
Let us point out that our CDMFT computations also include a Hartree-Fock component for the interaction terms that span neighboring clusters: correlations are only treated exactly within the cluster (here, the unit cell).

Leaving technical details of the static HF-MFT calculation to Appendix.~\ref{sec:MF}, let us only discuss the main results. 
They are presented in Fig.~\ref{fig:compr} for one of the optimal interaction parameter sets of Fig.~\ref{fig:dos}\itx{C} and discussed in Sec.~\itx{\ref{sec:Vpd}}.  
The HF-MFT results for the nematic order parameter (red curves in Fig.~\ref{fig:compr}, for MFT) show similar behavior and magnitude as in CDMFT (open cyan symbols) from low to intermediate doping $x$.
At very large doping, however, the CDMFT computation runs into convergence issues, most likely due to sector changes \cite{dionne:2023}, frequent to ED-CDMFT. 
The static HF-MFT data, on the other hand, extends until the unit-cell becomes empty of electrons. 
As expected, the order parameter ($\eta$) in the MFT goes back down at large doping, as discussed in Sec.~\itx{\ref{role_Vpp}}.

This case illustrates perfectly our point about the importance of computing not only charge nematicity, which is essentially the same in both methods here, but also the DOS. 
The lower part of Fig.~\ref{fig:compr} shows the Hartree-Fock mean-field DOS. 
It does not have an UHB nor, equivalently, a CTG. 
Of course, the crucial difference between CDMFT and MFT lies in the dynamics.
The density of states computed in CDMFT shows a splitting  of correlated bands due to the local interaction, in particular the existence of a UHB (see Fig.~\ref{fig:dos}).
Given the dynamical nature of the probe used to detect nematicity~\cite{wang2023_nema} a dynamical method like CDMFT is indicated to study the problem theoretically.

\section{Summary and Conclusion}\label{Sec:Discussion}

In this paper, we analyzed the three-band Hubbard (or Emery) model to investigate the interaction-induced intra-unit-cell (IUC) charge nematicity in hole-doped cuprates. 

Cluster dynamical mean-field theory (CDMFT) allowed us to solve the model in a regime where neither electronic structure nor interactions dominate. 
In addition, with this method, densities of states exhibit an upper Hubbard band in agreement with experimental observation when the values of $U_d$ and $U_p$  satisfy the minimum requirement $U_d \gg U_p$. 
We also argued that the charge nematicity by itself is not enough to claim agreement with experiments since different methods can give the same charge nematicity but unphysical densities of states.  

We first looked for model parameters that could lead to small nematicity at low doping and found that it was important to study the interplay of two different nearest-neighbor interactions: $V_{pp}$ (between $p_x$ and $p_y$ electrons of planar oxygens) and $V_{pd}$ (between $p_{x,y}$ electrons of planar oxygens and $d_{x^2-y^2}$ electron of copper).
As expected, increasing $V_{pp}$ shifts the onset of the nematic transition to low doping, where it is observed experimentally. (see Fig.~\ref{fig:phase}\itx{A} and top panel of Fig.~\ref{fig:phase}\itx{C}). 
However, $V_{pd}$ competes with $V_{pp}$ and pushes the onset of the transition to high doping for large  $V_{pd}$ (see Fig.~\ref{fig:phase}\itx{B} and bottom panel of Fig.~\ref{fig:phase}\itx{C}). 

$V_{pd}$ cannot be neglected: The requirements of (i) small nematicity at low doping and (ii) well-resolved ZRSB and UHB (see Fig.~\ref{fig:dos}\itx{A}), force us to conclude that intermediate values of $V_{pd}$ are optimal to obtain both a DOS compatible with observations and small nematicity at low doping.

Contact with experiment is achieved by computing the splitting of energy levels of the $p_x$ and $p_y$ orbitals of oxygen atoms for the optimal interaction parameter set, as shown in Fig.~\ref{fig:dos}\itx{C}. 
The UHB is not much affected by nematicity and the CTB energy spliting ($\delta\ve$) between the $p_x$ and $p_y$ partial DOS shadows the splitting observed in the ZRSB. 
The calculated $\delta\ve$ semi-quantitatively agrees with the recently measured value in Bi$_2$Sr$_2$CaCu$_2$O$_{8+x}$ using sublattice-resolved spectroscopic imaging scanning tunneling microscope \cite{wang2023_nema}, see Sec.~\itx{\ref{sec:extract}} for details. Note that the ZRSB is not well resolved in these experiments.
 
Given the available experimental observations there is still, however, some uncertainty in the precise value of nearest-neighbor interactions $V_{pp}$ and $V_{pd}$.
Since many of the properties of cuprates can be explained without taking into account these interactions, we argue that their effect is generally small and that more accurate values may not be needed until warranted by additional observations. 
For most purposes, the dominant interaction remains $U_d$ the on-site repulsion on copper orbitals.

\begin{acknowledgments}
We thank B. B-Labreuil and N. Martin for useful discussions. A.K. acknowledges support from Canada First Research Excellence Fund and by the Natural Sciences and Engineering Research Council of Canada (NSERC) under Grant No. RGPIN-2019-05312. 
Computing resources were provided by Calcul Qu\'ebec and by the Digital Research Alliance of Canada.
\end{acknowledgments}

\appendix
\section{Methods}\label{Sec:methods}

Cluster dynamical mean-field theory (CDMFT) \cite{kotliar:2001, maier:rmp, KotliarRMP:2006, LTP:2006, senechal:2012, dionne:2023, senechal:intro} is a cluster extension of DMFT \cite{georges:rmp}. 
In this approach, instead of a single-site impurity, we consider a cluster of multiple sites as an impurity with open boundary conditions, taking into account short-ranged spatial correlations exactly. The effect of the cluster's environment is accounted for by a set of additional, uncorrelated ``bath'' orbitals hybridized with the cluster. 
The bath orbitals have their own energy levels $\epsilon_{i\sigma}$, which may or may not be spin dependent, and are hybridized with the cluster sites (labeled $r$) with amplitudes $\theta_{ir\sigma}$.
The lattice is tiled with these clusters and the parameters of the bath orbitals are set by requiring some consistency between the local (to the cluster) electron Green function and the impurity model Green function.

We used a three-site cluster (Cu, O$_x$ and O$_y$), with each site connected with two bath orbitals, as shown in Fig.~\ref{fig:comb}\itx{A}(inset). 
Here the cluster coincides with the unit cell of the model.
The six bath orbitals represent the effect of the rest of the lattice.
For simplicity, each bath orbital is hybridized to one physical orbital only.
Hence the parameters of the impurity model include six bath energy levels ($\epsilon_{1...6}$) and six hybridization parameters ($\theta_{1...6}$), all to be determined self-consistently. 

In the cluster-bath model, Fig.~\ref{fig:comb}\itx{A}(inset), nematicity is naturally allowed since all bath energies ($\epsilon_i$) and hybridization parameters ($\theta_i$) are independent. 
The normal-state solution can be imposed by setting $\epsilon_{3,4} = \epsilon_{5,6}$ and $\theta_{3,4} = \theta_{5,6}$ (one-to-one correspondence). This constraint was imposed in order to obtain the normal-state solutions (open symbols) in Fig.~\ref{fig:comb}\itx{C}. 

At every CDMFT iteration \cite{senechal:2012, senechal:intro}, the impurity electron Green function $G_c(\omega)$ is obtained by solving the impurity problem with an exact diagonalization solver at zero temperature. The electron self-energy is then extracted from 
$G_c(\omega)$ via Dyson's equation and the lattice electron Green function $G(\kv, \omega)$ is assembled from this self-energy and the exact electron dispersion $\epsilon(\kv)$. The local Green function $\bar G(\omega)$ is computed from $G(\kv, \omega)$ by Fourier transforming back to the cluster, and the difference $|\bar G^{-1}(\omega)-G^{-1}_c(\omega)|$ is minimized over the set of bath parameters. This new set of bath parameters is then used for the next iterations.
The on-site interactions ($U_p$ and $U_d$) and the extended interactions ($V_{pp}$ and $V_{pd}$) within the cluster are treated exactly, whereas the extended interactions between clusters are treated with Hartree-Fock mean-field theory.
Converging the mean-field procedure is done simultaneously as converging the DMFT procedure, over the same iterations.
This procedure was also used  recently in Ref.~\cite{kundu:2023}.
In this work it takes on average 40 iterations to achieve self-consistency.
Once the procedure has converged, the lattice and cluster averages of one-body operator, $\hat{O}=\sum_{\alpha, \beta} O_{\alpha\beta} c_\alpha^\dagger c_\beta$, can be calculated with the help of the converged Green function:
\beq
\label{lavg}
\langle \hat{O} \rangle = \oint \frac{d\omega}{2\pi} \frac{d^2\bk}{(2\pi)^2} \text{Tr} \left[\mathbf{O}(\bk) G(\bk, \omega) \right],
\eeq
\beq
\label{cavg}
\langle \hat{O} \rangle_c = \oint \frac{d\omega}{2\pi} \text{Tr} \left[\mathbf{O} G_c(\omega) \right].
\eeq
We used Eq.~\ref{lavg} to calculate partial orbital densities and the nematic order parameter. 

Finally, a few remarks on our choice of cluster model for the CDMFT computation. 
We used a 3-site cluster [Fig.~\ref{fig:comb}\itx{A}(inset)], which is of same size as the CuO$_2$ unit-cell [Fig.~\ref{fig:comb}\itx{A}(main panel)].
This choice of cluster allows us to take into account the effect of $V_{pp}$ and $V_{pd}$ beyond mean field, although a substantial fraction of these interactions (between neighboring clusters) must be treated at the Hartree-Fock level.
A different cluster system was adopted in Ref.~\cite{AMT:pnas} to study superconductivity in cuprates: a Cu-cluster consisting of 4 Cu atoms, and an O-cluster consisting of 8 O atoms.
But in that work only $U_d$ was considered; if the effect of other interaction terms ($U_p$, $V_{pp}$, $V_{pd}$) is to be included beyond mean-field theory, then that cluster system is no longer adequate.

\section{Hartree-Fock approximation}
\label{sec:MF}

In this appendix, we provide technical details on the Hartree-Fock mean-field theory (HF-MFT) and derive the three-band MF Hamiltonian to calculate the order parameter $\eta$ discussed in Sec.~\ref{sec:comparison} and shown in Fig.~\ref{fig:compr}. 
We also provide more CDMFT data in order to compare them with the HF-MFT results. We have argued in the MT (see Sec.~\ref{sec:comparison}) that the DOS from HF-MFT is in strong disagreement with experiment. The value of the order parameter is insufficient to judge the accuracy of the theory. 

We derive the mean field Hamiltonian (within Hartree and Fock approximations), which, in the momentum space can be written as
\beq
\label{Ham:MF}
\mathcal{H}_\text{MF} = \sum_{\bk, s} \Psi_{\bk, s}^\dagger \mathcal{H}_{\bk, s} \Psi_{\bk, s} + f(n_{p_x}, n_{p_y}, n_d),
\eeq
where 
\beq
\label{MF}
\mathcal{H}_{\bk, s} =
\begin{pmatrix}
\xi_x & \gamma_2(\bk) & \gamma_1(k_x) \\
\tilde{\gamma}_2(\bk) & \xi_y & \gamma_1(k_y) \\
\tilde{\gamma}_1(k_x) & \tilde{\gamma}_1(k_y) & \xi_d
\end{pmatrix},
\eeq
with the multiplet $\Psi_{\bk s}^\dagger = (p_{x\bk s}^\dagger, p_{y\bk s}^\dagger, d_{\bk, s}^\dagger)$ containing the creation operators for electrons on oxygen $p_x$ and $p_y$ orbitals and copper $d_{x^2-y^2}$ and with $f(n_{p_x}, n_{p_y}, n_d, n_{p_x d}, n_{p_x p_y})$ defined by 
\beq
\label{energy}
\begin{split}
&\frac{f(n_{p_x}, n_{p_y}, n_d, n_{p_x d}, n_{p_x p_y})}{N} \\
&\hspace{0.1cm} = -\tilde{U}_p \frac{(n_{p_x} + n_{p_y})^2}{8} + \tilde{V}_{pp} \frac{(n_{p_x} - n_{p_y})^2}{8} - \tilde{U}_d \frac{n_d^2}{4} \\
&\hspace{0.5cm} + V_{pd} \left( n_{p_x d} n_{d p_x} + n_{p_y d} n_{d p_y} \right)
+ 2V_{pp} n_{p_x p_y} n_{p_y p_x}.
\end{split}
\eeq
and
\beq
\label{renorm}
\begin{split}
\tilde{U}_p &= U_p + 8V_{pp} - 8 V_{pd}, \\
\tilde{V}_{pp} &= 8V_{pp} - U_p, \\
\tilde{U}_d &= U_d - 4 V_{pd}.
\end{split}
\eeq
In ~\er{energy}, $n_{p_{x(y)}}$ is the density of $p_{x(y)}$ electrons, $n_d$ is the density of $d_{x^2-y^2}$ electrons,
\beq
n_{p_\nu d} n_{dp_\nu} \equiv (1/N)\sum_{i, s} \sum_\nu \langle p_{i+\nu/2,s}^\dagger d_{i,s} \rangle \langle d_{i,s}^\dagger p_{i+\nu/2,s} \rangle
\eeq
and
\beq
n_{p_x p_y} n_{p_y p_x} \equiv
\frac1{4N}\sum_{i, s} \sum_{\nu\nu'} \langle p_{i+\nu/2,s}^\dagger p_{i+\nu'/2,s} \rangle
\langle p_{i+\nu'/2,s}^\dagger p_{i+\nu/2,s} \rangle~~,
\eeq
with $s$ the spin index. 
Here, $n_{p_\nu d} = n_{d p_\nu}$ and $n_{p_x p_y} = n_{p_y p_x}$, due to hermiticity of the mean-field Hamiltonian \er{Ham:MF}.
The cross-densities in ~\er{energy}, $n_{p_\nu d}$ and $n_{p_\nu p_{\nu'}}$, originate from the Fock approximation, which also renormalizes hopping elements (off-diagonal terms) in ~\er{MF}:
\beq
\label{hopp}
\begin{split}
\gamma_1(k_i) &= 2 \left( t_{pd} - \frac{V_{pd}}{2} n_{dp_i} \right) \cos(k_i/2), \\
\tilde{\gamma}_1(k_i) = \gamma_1^\dagger(k_i) &= 2 \left( t_{pd} - \frac{V_{pd}}{2} n_{p_id} \right) \cos(k_i/2), \\
\gamma_2(\bk) &= 4 \left( t_{pp} - \frac{V_{pp}}{2} n_{p_y p_x} \right) \cos(k_x/2) \cos(k_y/2), \\
\tilde{\gamma}_2(\bk) = \gamma_2^\dagger(\bk) &= 4 \left( t_{pp} - \frac{V_{pp}}{2} n_{p_x p_y} \right) \cos(k_x/2) \cos(k_y/2).
\end{split}
\eeq
Finally, the diagonal elements of ~\er{MF} can be written as
\beq
\begin{split}
\xi_x &= \Delta + \tilde{U}_p \frac{n^p_x + n^p_y}{4} - \tilde{V}_{pp} \frac{n^p_x - n^p_y}{4} - \mu, \\
\xi_y &= \Delta + \tilde{U}_p \frac{n^p_x + n^p_y}{4} + \tilde{V}_{pp} \frac{n^p_x - n^p_y}{4} - \mu, \\
\xi_d &= \tilde{U}_d \frac{n^d}{2} - \mu,
\end{split}
\eeq
with $\tilde{U}_d$, $\tilde{U}_p$ and $\tilde{V}_{pp}$ given in ~\er{renorm},
which has interaction-dependent terms arising due to the Hartree approximation only. 
This is expected because Hartree terms are local hence they renormalize on-site terms of the Hamiltonian, not the hopping terms; the latter are renormalized by Fock terms \er{hopp}.

For given values of $n_{p_x}$, $n_{p_y}$ and $n_d$, the chemical potential ($\mu$) is implicitly given by solving
\beq
\frac{n_{p_x} + n_{p_y} + n_d}{3} \equiv n = \frac{1}{N} \sum_{\alpha, \bk, s} \frac{1}{e^{\omega_{\alpha, \bk, s}/T} + 1},
\eeq
where $\omega_{\alpha, \bk, s}$, with $\alpha=1,2,3$ as the band index, is the dispersion of three bands with mixed orbital character, obtained upon diagonalizing the mean-field Hamiltonian \er{Ham:MF}. 
w
The values of orbital densities $n_{p_x}$, $n_{p_y}$ and $n_d$, and cross-densities, $n_{p_x d}$, $n_{p_y d}$ and $n_{p_x p_y}$, are obtained self-consistently. The nematic order parameter is obtained from the density imbalance of $p_x$ and $p_y$ electrons, as defined in ~\er{nem_order} and discussed in detail in Sec.~\ref{sec:order}. We quote results of the Hartree-Fock approximation since they are quite different from those of the Hartree approximation and they should be better.

The HF-MFT results for band parameters as given by \er{param} of the MT are shown by red curves in the top panel of Fig.~\ref{fig:compr} of the MT for a given set of interaction parameters as mentioned in the figure itself. 
In the same figure, the cyan curves are results obtained from CDMFT, whose DOS at hole doping $x=0.2$, which is close to the transition point approaching from the ordered phase, is shown in Fig.~\ref{fig:dos}\itx{C} of the MT. 
As explained in the MT, despite the similarity between MFT and CDMFT for the value of the order parameter, the DOS from HF-MFT is unsatisfactory compared with the experiment (see Fig.~\ref{fig:compr}(Bottom panel), which does not show upper Hubbard band), so we must reject it. 

\begin{figure}
\center{\includegraphics[width=0.9\linewidth]{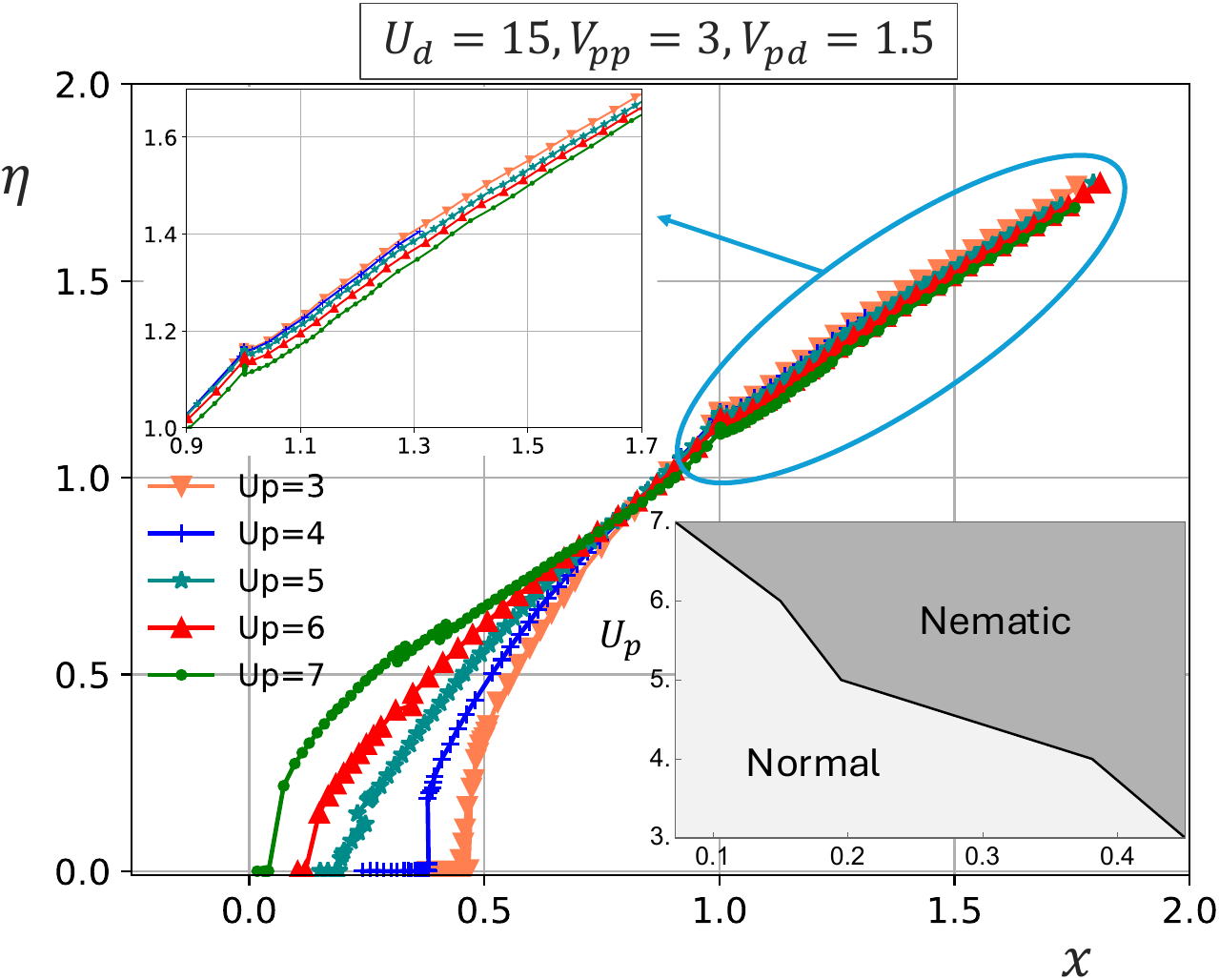}}
\caption{Nematic order parameter ($\eta$) as a function of hole doping ($x$) exploring the role of $U_p$. At low doping, $U_p$ is found to favor nematicity since the onset of nematic transition shifts near half-filling as $U_p$ increases. Inset (top): The feature in the high doping region is blown up for clarity. Inset (bottom): The phase diagram demonstrating that large $U_p$, or strongly correlated O orbitals, is favorable to charge nematicity.  
}\label{fig:Up_SI}
\end{figure}

\begin{figure}
\center{\includegraphics[width=0.8\linewidth]{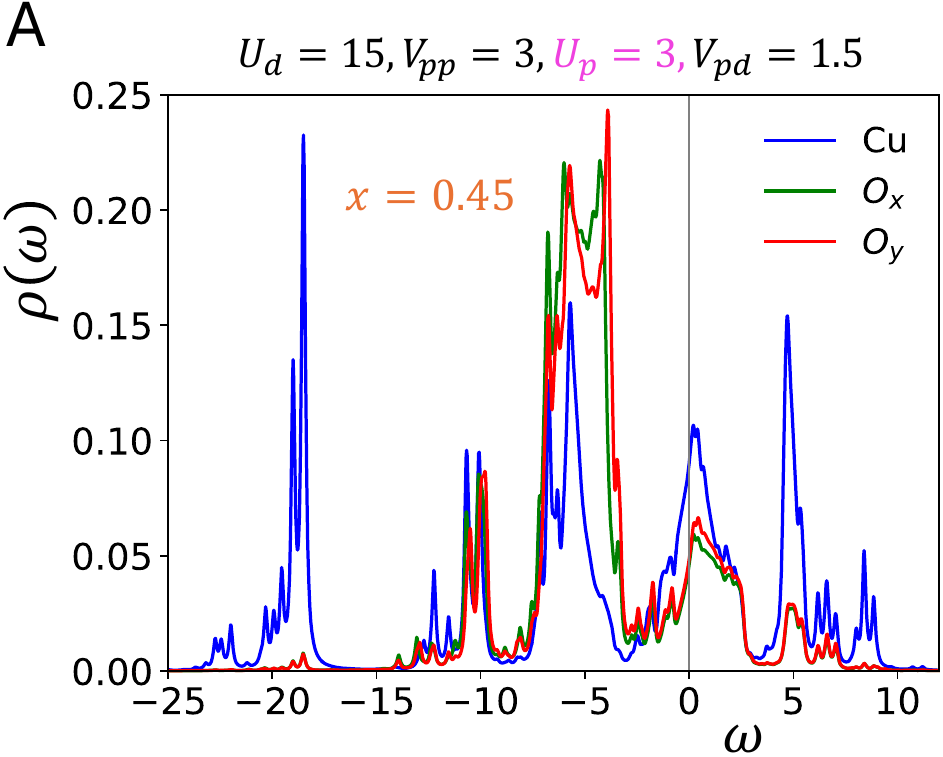}}
\center{\includegraphics[width=0.8\linewidth]{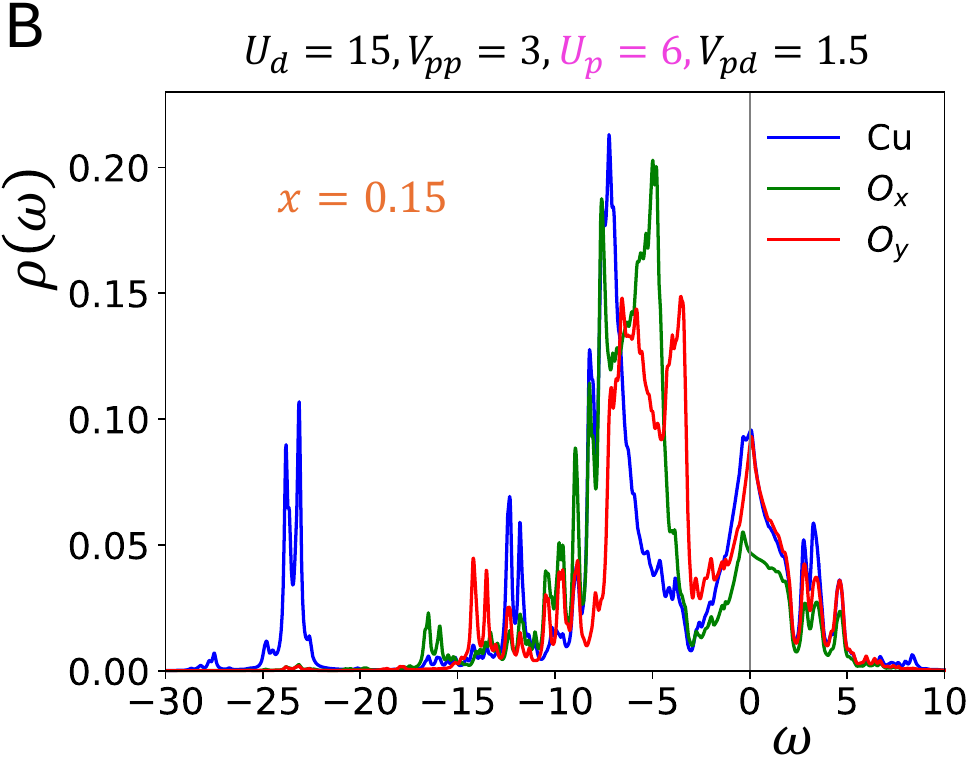}}
\caption{The figure illustrates that increasing $U_p$ from $U_p=3$ in (\itx{A}) to $U_p=6$ in (\itx{B}) leads to an increase in hybridization and consequently to an increase in nematicity. 
However, it is not physically acceptable because the UHB gets absorbed by the ZRSB. 
The doping ($x$) is not the same in both figures because at the value of $x$ considered in (\itx{B}) to obtain DOS for $U_p=6$, i.e., $x=0.15$, the system is non-nematic for $U_p=3$ in (\itx{A}); see orange curves in Fig.~\ref{fig:Up_SI}. 
The nematic order parameter develops only at larger $x$; for (\itx{A}), we choose $x=0.45$, which is closest to the transition approaching from the ordered side, as can be also seen in Fig.~\ref{fig:Up_SI}. 
\label{fig:dos_Up} }
\end{figure}

\section{Role of $U_p$ on charge nematicity}
\label{sec:Up}
In this appendix we discuss the influence of $U_p$ on the physics. We show that both HF-MFT and CDMFT predict similar behavior of $\eta$ vs hole doping ($x$) as a function of $U_p$.
DOS plots show that at fixed $V_{pd}$, even though higher Up favors nematicity by shifting the onset of the transition closer to half-filling, the DOS contradicts experiments.

The origin of the increase in nematicity from low to intermediate doping upon increasing $U_p$ was discussed in the MT. 
The CDMFT data supporting that argument is provided in Fig.~\ref{fig:Up_SI}. 
The raw CDMFT data is provided in the main panel, while the nematic transition in the parameter space is summarized in the phase diagram (lower inset). 
The phase diagram suggests that the onset of the transition shifts to lower doping as $U_p$ increases, giving rise to the increase in the order parameter ($\eta$).
The upper inset is the blown up part of $\eta$ at large doping. 
Although $\eta$ seems to decrease in this doping region as $U_p$ increases, the decrease is too small to conclude anything. 
Anyway, since it is possible that at this large doping the band parameters of the model would be massively different if they are derived from DFT, we do not dwell on it further.

In the DOS plots, as illustrated in Fig.~\ref{fig:dos_Up}, increasing $U_p$ leads to an increased hybridization where the UHB is absorbed by the ZRSB in the low doping region of interest, as seen for example in Fig.~\ref{fig:dos_Up}\itx{B}. 
At the same $V_{pd}$, for too small $U_p$, however, although the ZRSB and the UHB are well resolved as shown in Fig.~\ref{fig:dos_Up}\itx{A} (obtained at rather large doping, $x=0.45$, in the nematic phase. At the doping same as in Fig.~\ref{fig:dos_Up}\itx{B}, the system is in the normal phase), the onset of the nematic transition is pushed at larger doping, see orange curves in Fig.~\ref{fig:Up_SI}. 
To obtain the optimal situation for $U_p=3$, we then need to decrease $V_{pd}$, which we will discuss in the next section.
Considering same $V_{pd}$ as in Fig.~\ref{fig:dos_SI}\itx{B}, it is, therefore, crucial to restrict ourselves to small enough, or intermediate, values of $U_p$ to get the results in agreement with the experiment. 

\begin{figure}
\center{\includegraphics[width=0.8\linewidth]{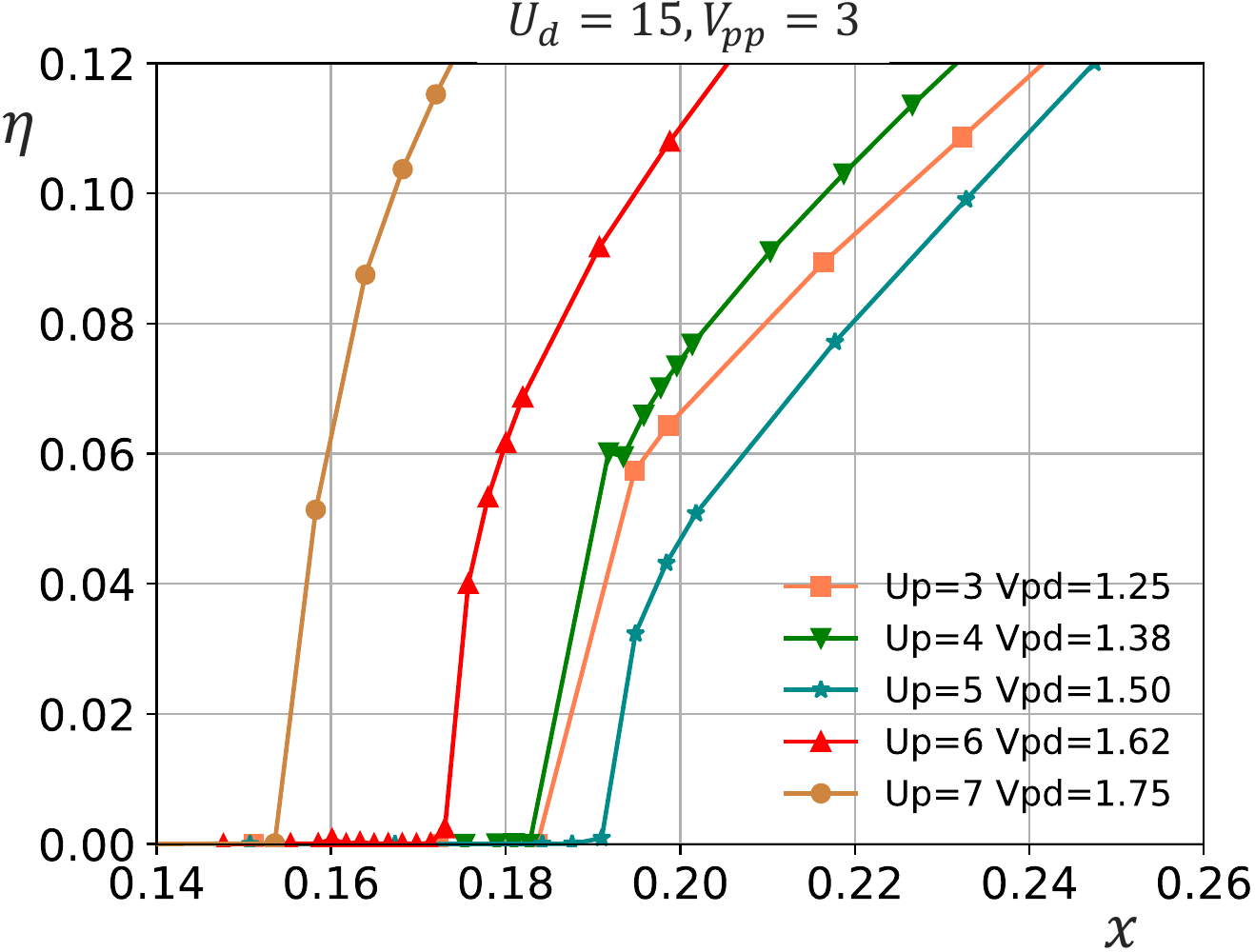}}
\caption{Nematic order parameter ($\eta$) as a function of hole doping ($x$) for various optimal interaction parameter sets. The weak nematicity at low doping is accompanied by well-resolved ZRSB and UHB with weak splitting of oxygen $p_x$ and $p_y$ orbitals near the Fermi level. The latter is shown explicitly in Fig.~\ref{fig:dos_SI} for few choices of optimal doping and interaction parameter sets.  
}\label{fig:optimal}
\end{figure}

\section{Other optimal sets of parameters}
\label{sec:optimal}

In this appendix we argue that if we do not restrict ourselves to small $U_p$, it is possible to change $V_{pd}$,  as discussed in Sec.~\ref{sec:Vpd}, to find reasonable CDMFT results for other sets of parameters than those we argued for in the main text. 

\begin{figure*}
\center{\includegraphics[width=0.80\linewidth]{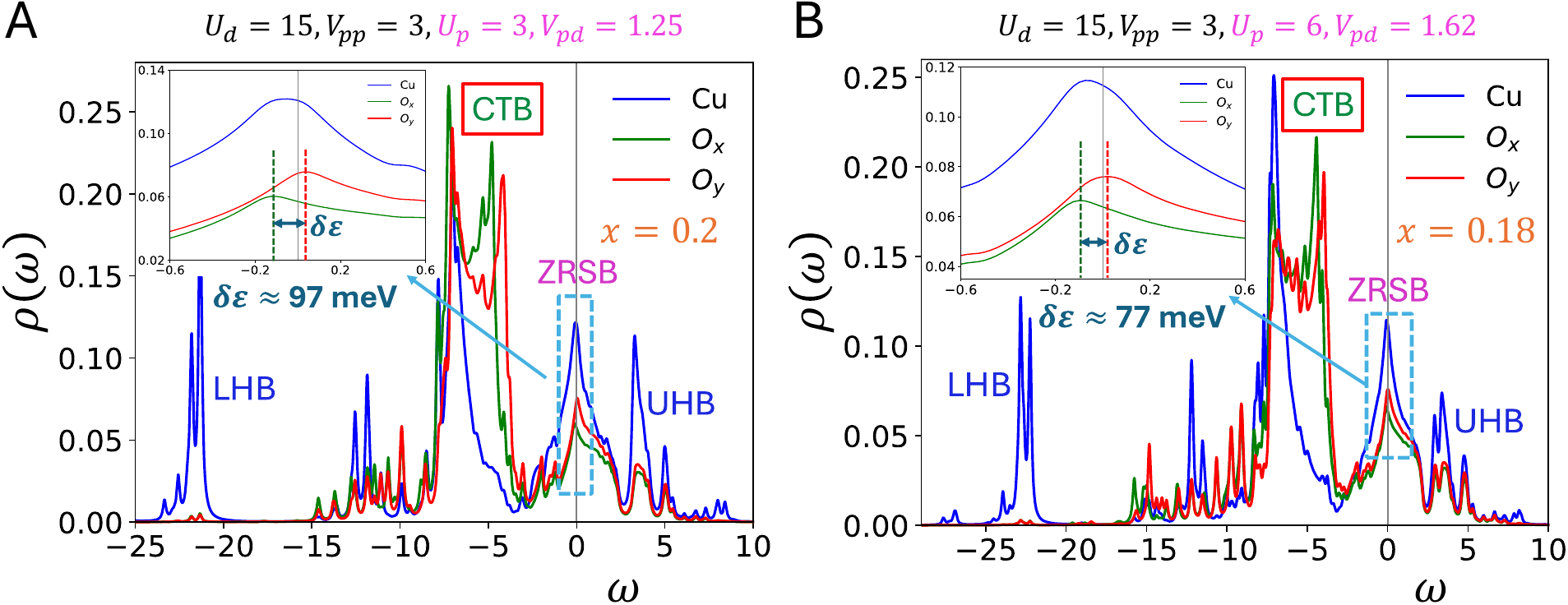}}
\caption{\label{fig:dos_SI} 
DOS at other optimum interaction parameter sets (low and high $U_p$). (\textit{A}) Low $U_p$ disfavors nematicity, so for $V_{pp}$ fixed at the same value as in Fig.~\ref{fig:dos}, the $V_{pd}$ must be optimized to get (i) the onset of nematic transition at low doping and, (ii) the ZRSB and the UHB being well resolved in the DOS. At $U_p=3$, the optimum scenario happens to be for $V_{pd}=1.25$. As demonstrated in Fig.~\ref{fig:dos}, for $V_{pd}$ larger than this value, the onset of the nematic transition is pushed at large doping, while for $V_{pd}$ smaller than this, the ZRSB and the UHB shows strong hybridization. 
(\textit{B}) Large $U_p$ favors nematicity. So for $U_p=6$, the parameter $V_{pd}=1.25$ as in panel (\textit{A}) is too small to resolve ZRSB and UHB. For the optimum situation,  $V_{pd}$ has to be increased to $V_{pd}=1.62$.}
\end{figure*}

When we leave ourselves with complete freedom in the choice of parameters, it is possible to obtain comparable results for the order parameter as a function of hole doping. 
In Fig.~\ref{fig:optimal}, we provide CDMFT data of $\eta$ vs $x$ for some of the other possible optimal parameter sets, which yields small nematicity at low doping and the UHB and ZRSB in the DOS are well resolved. 
The DOS plots for few choices of optimal interaction parameters sets are shown in Fig.~\ref{fig:dos_SI}.
At any rate, limiting ourselves to small-enough $U_p$ helps to restrict the range of acceptable interaction parameters. 
So, we chose $U_p=5$ in the MT to discuss all our findings.

%

\end{document}